\documentclass[10pt,twocolumn,aps,floatfix,citeautoscript,superscriptaddress,prl]{revtex4-2}
\usepackage[english]{babel}

\usepackage{graphicx}
\usepackage{subfig}
\usepackage{natbib}
\usepackage{physics}
\usepackage{tikz}
\usepackage{xcolor}

\usepackage{amsmath, amssymb}  
\usepackage{multirow}       
\usepackage{xcolor}         
\usepackage{pgfplots}       
\usepackage{ulem}           
\usepackage{listings}       
\usepackage{amsfonts}       
\usepackage{comment}        
\usepackage{float}          
\usepackage[colorlinks=true, allcolors=blue]{hyperref}       

\definecolor{codegreen}{rgb}{0,0.6,0}
\definecolor{codegray}{rgb}{0.5,0.5,0.5}
\definecolor{codepurple}{rgb}{0.58,0,0.82}
\definecolor{backcolour}{rgb}{0.95,0.95,0.92}

\newcommand{\qea}{\alpha^{BO}}

\definecolor{graphcol}{HTML}{44AA99}

\graphicspath{{./}{Figures/}}

\makeatletter
\long\def\@makecaption#1#2{%
  \vskip\abovecaptionskip
  \begingroup
    \small
    \parindent=0pt
    \leftskip=0pt
    \rightskip=0pt plus 1fil
    \parfillskip=0pt
    \textbf{#1:} #2\par
  \endgroup
  \vskip\belowcaptionskip
}
\makeatother

\begin{document}

\title{Many-body cages: disorder-free glassiness from \\ flat bands in Fock space, and many-body Rabi oscillations}

\date\today
\author{Tom Ben-Ami}
\affiliation{Theoretical Physics III, Center for Electronic Correlations and Magnetism, Institute of Physics, University of Augsburg, D-86135 Augsburg, Germany}
\affiliation{Max-Planck-Institut f\"{u}r Physik komplexer Systeme, N\"{o}thnitzer Stra\ss e 38, Dresden 01187, Germany}

\author{Markus Heyl}
\affiliation{Theoretical Physics III, Center for Electronic Correlations and Magnetism, Institute of Physics, University of Augsburg, D-86135 Augsburg, Germany}
\affiliation{Centre for Advanced Analytics and Predictive Sciences (CAAPS), University of Augsburg, Universitätsstr. 12a, 86159 Augsburg, Germany}

\author{Roderich Moessner}
\affiliation{Max-Planck-Institut f\"{u}r Physik komplexer Systeme, N\"{o}thnitzer Stra\ss e 38, Dresden 01187, Germany}

\begin{abstract}
We introduce many-body caging as a novel mechanism for nonthermal behaviour in quantum matter. We define many-body cages as eigenstates that, through quantum interference, become localised on a subgraph of the many-body state graph. These many-body cages can lead to the formation of flat bands in the many-body spectrum at characteristic, system-independent energies. These flat
bands can realize a novel type of glassy eigenspectrum order in the absence of disorder, which we quantify by a band-overlap order parameter with an intricate, possibly fractal, distribution over the many-body state graph. We further show that these many-body cages exhibit distinctive signatures in experimentally accessible quantities, such as through a nonvanishing long-time memory of the initial condition, and many-body Rabi oscillations set by the characteristic flat band energies. While our predictions in principle apply to any constrained quantum system, we demonstrate them here for 2D lattice gauge theories and models relevant for current experiments in Rydberg atoms. We expect that these many-body cages offer a promising route to realize nonequilibrium quantum states with novel properties.
\end{abstract}

\maketitle
\textit{Introduction --}
Over the past two decades, quantum simulation experiments have established the nonequilibrium real-time dynamics of isolated 2D quantum many-body systems as a central frontier in quantum physics. Driven by developments across diverse platforms, ranging from Rydberg atomic systems~\cite{rydbergs}, ultracold quantum gases~\cite{ultracoldatoms}, trapped ions~\cite{trappedions}, to superconducting quantum processors~\cite{squbits}, these experiments provide access to experimental observations that were previously beyond reach.
At long times, coherent quantum dynamics typically gives rise to thermodynamic behaviour, as described by the eigenstate thermalisation hypothesis~(ETH)~\cite{deutsch1991quantum,srednicki1994chaos,d2016quantum}. This raises a fundamental question: can quantum matter exhibit stable nonequilibrium phases with emergent properties that go beyond conventional statistical ensembles?

Many-body localisation (MBL) offers a paradigm for such behaviour~\cite{basko2006metal, abanin2019colloquium, alet2018many, nandkishore2015many}, primarily in one dimension~\cite{serbyn2021quantum, turner2018weak, chandran2023quantum}, where it supports nonequilibrium phases such as spin glasses and discrete time crystals~\cite{2018RPPh...81a6401S, RevModPhys.95.031001,yao2018time,2019arXiv191010745K}, characterised by eigenstate order and robustness against thermalisation.
However, identifying generic mechanisms which evade thermalisation and host nonequilibrium physics for $d>1$ remains an open and exciting frontier.

In this work, we introduce many-body cages as a new, {\it disorder-free} route to nonequilibrium many-body quantum dynamics.
Many-body cages are eigenstates localised on a subgraph of the many-body state graph through destructive quantum interference.
They are inspired by flat-band localisation in single-particle spectra, where interference and lattice topology gives rise to compact localised states (CLS)~\cite{sutherland1986localization, kirkpatrick1972localized, zhang2020compact, hart2020compact}, and to Aharonov-Bohm cages in single-particle systems~\cite{vidal1998aharonov,2025NatPh..21..221C}. We extend this mechanism to many-body systems by showing that the local topology of the many-body state graph can give rise to CLS and degenerate flat bands in the many-body spectrum.
Many-body cages constitute a mechanism for nonthermal behaviour different from other known ones such as Hilbert-space fragmentation, many-body scars, and disorder-free localisation, as we describe in the discussion.

Concretely, across a broad class of constrained quantum systems, we identify many-body cages on the many-body state graph, and find that these form flat bands in the many-body spectrum at characteristic energies determined via a graph-theoretical construction.
These many-body flat bands host a novel form of disorder-free glassy order: a many-body-caged spin glass, characterised by a band-overlap order parameter which quantifies how nonthermal the system is due to many-body cages.
Remarkably, in some cases this order displays a structure reminiscent of the fractal devil’s staircase and the Farey sequence—a signature of emergent arithmetic complexity.

A particularly distinctive dynamical feature of many-body cages is the appearance of coherent, Rabi oscillations between distinct many-body states, with a frequency proportional to the energy spacing between flat bands. This provides a dynamical fingerprint of the underlying many-body state graph topology. 

\begin{figure*}[!htb]
\centering
\includegraphics[width=1.0\linewidth]{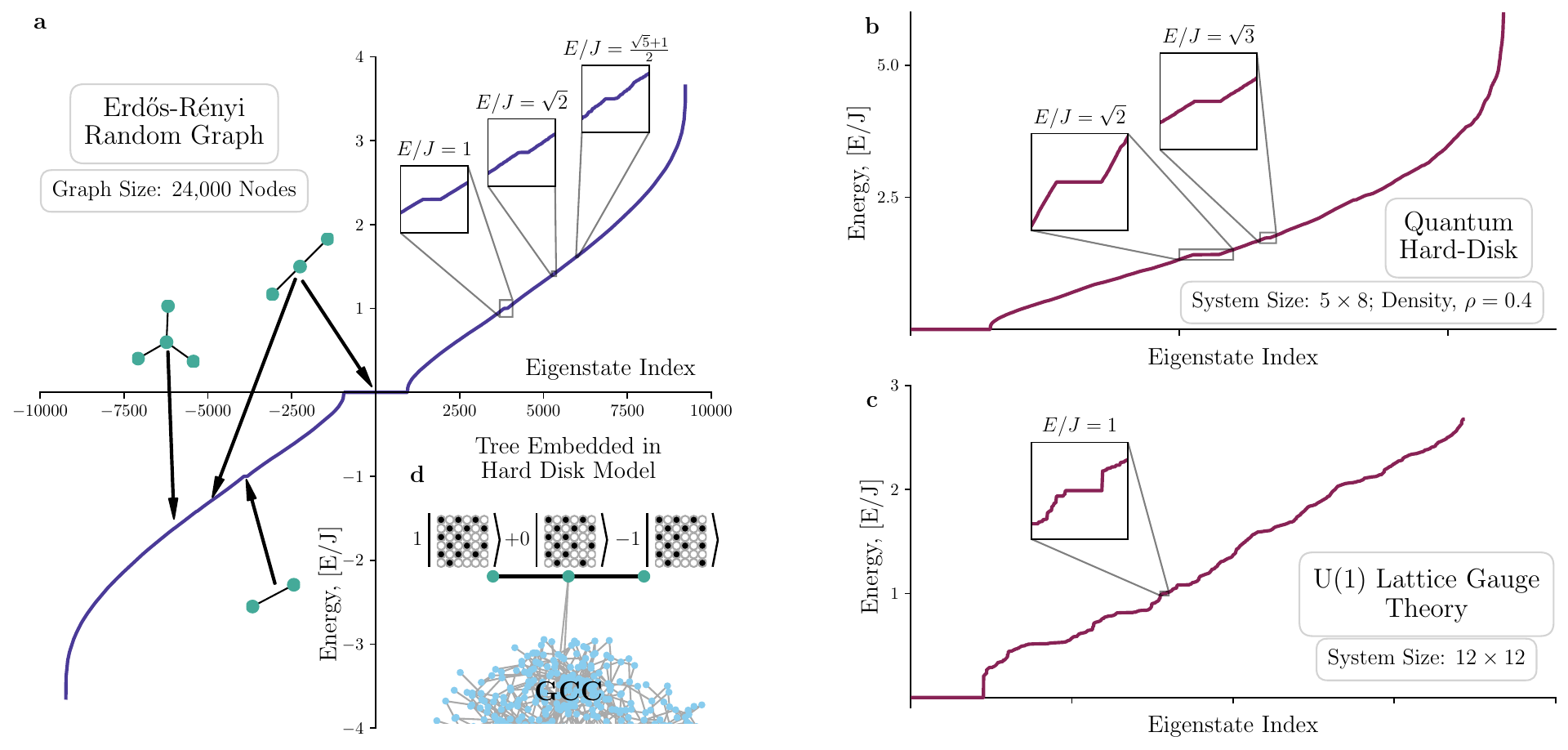}\:
\caption{\textbf{Many-body flat bands.} The Sparse Erd\H{o}s-R\'{e}nyi random graph \textbf{a}, state graph of the hard-disk model, \textbf{b}, and the spin-1/2 U(1) lattice gauge theory with a single dynamical charge \textbf{c}. For \textbf{b}\&\textbf{c} only half the spectrum is shown, as it is symmetric around $E=0$. Flat bands at specific closed-form values can be seen on all three spectra. Examples of tree structures associated with the flat-band energy eigenvalues are illustrated on the spectrum of the Erd\H{o}s-R\'{e}nyi random graph. \textbf{d} An example of a grafted tree of type $T_3$ in the hard-disk system.}
\label{fig:spec}
\end{figure*}

\textit{Many-body cages -- } 
To formalise many-body cages and flat bands, we consider the state graph of a quantum many-body system. The nodes of the state graph correspond to many-body basis states (e.g., a Fock state), and every off‑diagonal matrix element of the Hamiltonian is represented by an edge linking the two corresponding nodes. We restrict ourselves to Hamiltonians that contain only off‑diagonal terms; diagonal contributions, which appear as on‑site potentials (self‑loops) on the graph, are omitted (see SM for discussion). Consequently, the problem reduces to a hopping model on the state graph, whose spectrum coincides with that of the underlying microscopic many-body Hamiltonian.

We define many-body cages as states supported only on a subset of the many-body basis states, where they are localised as a result of destructive interference on the many-body state graph. In homogeneous systems, these cages lead to the formation of degenerate flat bands in the many-body spectrum at characteristic, system-independent energies determined by the eigenspectrum of the underlying local motifs.

The most striking manifestations of many‑body cages arise when a local constraint forbids certain many-body basis states. Familiar examples include hard‑disk, hard‑core dimer, and U(1) gauge models \citep{naik2024quantum,moessner2010quantum,chakraborty2025fractional}. Such constraints prune nodes from the state graph; when the pruning is not so severe as to lead to strong fragmentation~\cite{sala2020ergodicity, yang2020hilbert}, there still remains a connected component containing a finite fraction of all allowed many-body states, called the giant connected component (GCC).
When additional connected components of the state graph exist that have zero weight in the thermodynamic limit \citep{sala2020ergodicity,chandran2023quantum}, this is known as weak fragmentation~\citep{2022RPPh...85h6501M,sala2020ergodicity,khemani2020localization}. Alternatively, strong fragmentation occurs when there is no GCC---i.e., when all connected components of the state graph involve a measure-zero set of nodes in the thermodynamic limit. The transition from weak to strong fragmentation corresponds to a percolation transition~\cite{callaway2000network}. 

In the following, we focus on the GCC in the weak‑fragmentation regime, and neglect contributions from isolated clusters. For concreteness, we examine two constrained models in this regime. First, the quantum hard-disk model~\citep{naik2024quantum}, which describes particles with an excluded volume on a lattice, where the transition from weak to strong fragmentation is dictated by the particle density. Second, a class of U(1) lattice gauge theories with dynamical matter, recently shown to host tunable sub-diffusive transport~\citep{chakraborty2025fractional}. We expand on these two systems and include their Hamiltonians explicitly in the End Matter.
We find that the spectra of the GCC in both models display flat bands at a subset of energies,
\begin{equation}
    \frac{E}{J}\in\Bigl\{0,\pm1,\pm\sqrt{2},\pm\frac{\sqrt{5}\pm1}{2},\pm\sqrt{3},\dots\Bigr\},
    \label{eq:tree_energies}
\end{equation}
where \(J\) denotes the (uniform) hopping amplitude on the state graph (Fig.~\ref{fig:spec}b–c). 
While both models have a local constraint, they differ in that the hard-disk model has kinetic constraints, while the U(1) lattice gauge theory hosts gauge constraints. This highlights the generality of the phenomenon.
These energy values are generally modified by the inclusion of diagonal terms and/or non-uniform hopping amplitudes. The stability of these flat bands against such terms on the state graph is discussed in the SM.

In order to develop a general understanding for the emergence of many-body cages in these model systems we now turn to the more abstract concept of Erdős–Rényi random graphs, which share the key structural features with our model systems.
An Erdős-Rényi random graph is constructed by randomly connecting each pair of nodes with probability $p$, which sets the sparsity of the graph. Close to the percolation threshold, such graphs become increasingly locally tree like~\citep{bollobas1998random}. In the weak fragmentation regime, constrained systems show similar qualitative behaviour---sparse, non-uniform connectivity.

On such sparsely connected trees of the state graph, one can identify a class of exact CLS of the adjacency matrix of the GCC through the so-called tree‑grafting procedure~\citep{golinelli2003statistics, bauer2001random} (see, e.g., Fig. \ref{fig:spec}d): consider a finite tree \(T_i\), attached to a graph through a single node, \(r\). If \(T_i\) possesses an eigenvector \(V\) with eigenvalue \(\lambda\) whose amplitude vanishes on \(r\), then there exists an eigenvector of the GCC \(W\) with the same eigenvalue \(\lambda\) that has no support outside of \(T_i\).
Consequently, when multiple grafted trees sharing the same eigenvalue \(\lambda\) exist, a flat band at \(\lambda\) emerges (see Fig. \ref{fig:spec}a).
Note that we consider tree grafting only for the purpose of interpretation, and do not graft them explicitly; instead, the systems we consider naturally realize analogous grafts.

The condition for the eigenvector to have zero weight on node \(r\) is sufficient, but not necessary. As shown in Refs.~\cite{golinelli2003statistics, bauer2001random}, eigenvalues of finite trees can give rise to CLS even when the corresponding tree eigenvectors have nonzero weight at the attachment site, provided the embedding into the GCC permits a strictly localised solution.

Concretely, small trees have eigenvalues given by Eq.~(\ref{eq:tree_energies}) (see Table~\ref{tab:trees} for the exact tree eigenvalues), and their localised eigenvectors are decoupled from the rest of the graph \citep{sutherland1986localization,kirkpatrick1972localized}.

\begin{table}[ht]
    \centering
    \begin{tabular}{c|c|c|c}
        $\textbf{n}$ & \textbf{Tree} & \textbf{Graph} & \textbf{Spectrum} \\
        \hline
        $1$ & $T_1$ & 
        \begin{tikzpicture}[scale=0.4]
            \node[circle,draw,fill=graphcol,inner sep=2pt] (a) at (0,0) {};
        \end{tikzpicture} & 
        $\left\{ 0 \right\}$ \\
        \hline
        $2$ & $T_2$ & 
        \begin{tikzpicture}[scale=0.4]
            \node[circle,draw,fill=graphcol,inner sep=2pt] (a) at (0,0) {};
            \node[circle,draw,fill=graphcol,inner sep=2pt] (b) at (1,0) {};
            \draw (a) -- (b);
        \end{tikzpicture} & 
        $\left\{ +1, -1 \right\}$ \\
        \hline
        $3$ & $T_3$ & 
        \begin{tikzpicture}[scale=0.4]
            \node[circle,draw,fill=graphcol,inner sep=2pt] (a) at (0,0) {};
            \node[circle,draw,fill=graphcol,inner sep=2pt] (b) at (1,0) {};
            \node[circle,draw,fill=graphcol,inner sep=2pt] (c) at (2,0) {};
            \draw (a) -- (b) -- (c);
        \end{tikzpicture} & 
        $\left\{-\sqrt{2}, 0, \sqrt{2} \right\}$ \\
        \hline
        $4$ & $T_{4,1}$ & 
        \begin{tikzpicture}[scale=0.4, baseline={(current bounding box.center)}]
            \node at (1,0.8) {}; 
            \node[circle,draw,fill=graphcol,inner sep=2pt] (a) at (1,0) {};
            \node[circle,draw,fill=graphcol,inner sep=2pt] (b) at (0,0.5) {};
            \node[circle,draw,fill=graphcol,inner sep=2pt] (c) at (2,0.5) {};
            \node[circle,draw,fill=graphcol,inner sep=2pt] (d) at (1,-1) {};
            \draw (a) -- (b);
            \draw (a) -- (c);
            \draw (a) -- (d);
        \end{tikzpicture} & 
        $\left\{ -\sqrt{3}, 0, 0, \sqrt{3} \right\}$ \\
        $4$ & $T_{4,2}$ & 
        \begin{tikzpicture}[scale=0.4]
            \node[circle,draw,fill=graphcol,inner sep=2pt] (a) at (0,0) {};
            \node[circle,draw,fill=graphcol,inner sep=2pt] (b) at (1,0) {};
            \node[circle,draw,fill=graphcol,inner sep=2pt] (c) at (2,0) {};
            \node[circle,draw,fill=graphcol,inner sep=2pt] (d) at (3,0) {};
            \draw (a) -- (b);
            \draw (b) -- (c);
            \draw (c) -- (d);
        \end{tikzpicture} & 
        $\left\{ \pm {(\sqrt{5} - 1)}/{2}, \pm {(\sqrt{5} + 1)}/{2} \right\}$ \\
    \end{tabular}
    \caption{All unique trees of up to four nodes and their corresponding eigenenergies.}
    \label{tab:trees}
\end{table}

We next corroborate this picture in both microscopic models. The quantum hard‑disk model (with a corresponding spectrum displayed in Fig.~\ref{fig:spec}b) exhibits non-thermal dynamics solely due to quantum interference~\citep{naik2024quantum}. The U(1) lattice gauge theory with dynamical matter (spectrum in Fig.~\ref{fig:spec}c), shows distinct flat bands. In both cases the spectra display the flat bands predicted by the tree‑grafting picture.
All the results shown in the figures of this work have been obtained by means of exact diagonalization.

A closer inspection of the energy spectra of our considered models (Fig.~\ref{fig:spec}) reveals that specific flat bands may be suppressed or enhanced (i.e., an increase in the degeneracy) compared to the Erd\H{o}s-R\'{e}nyi model. In the Erd\H{o}s-R\'{e}nyi model, all tree structures are allowed, while physical constraints in microscopic models, in general, favour or exclude particular trees on the state graph. 
For instance, for quantum hard disks, the $E/J = \pm1$ bands are absent, while the $E /J= \pm\sqrt{2}$ bands are prominently enhanced (see Fig.\ref{fig:spec}b).

Many-body cages imply nonthermal behaviour: the component of a basis state supported on a many-body cage is localised by destructive interference, thereby violating ETH.
Furthermore, we present numerical evidence (see SM) that the degenerate flat bands span a finite fraction of Hilbert space, in both Erd\H{o}s-R\'{e}nyi graphs and the quantum hard-disk model.

\begin{figure}[!hbt] 
\centering
\includegraphics[width=0.95\linewidth]{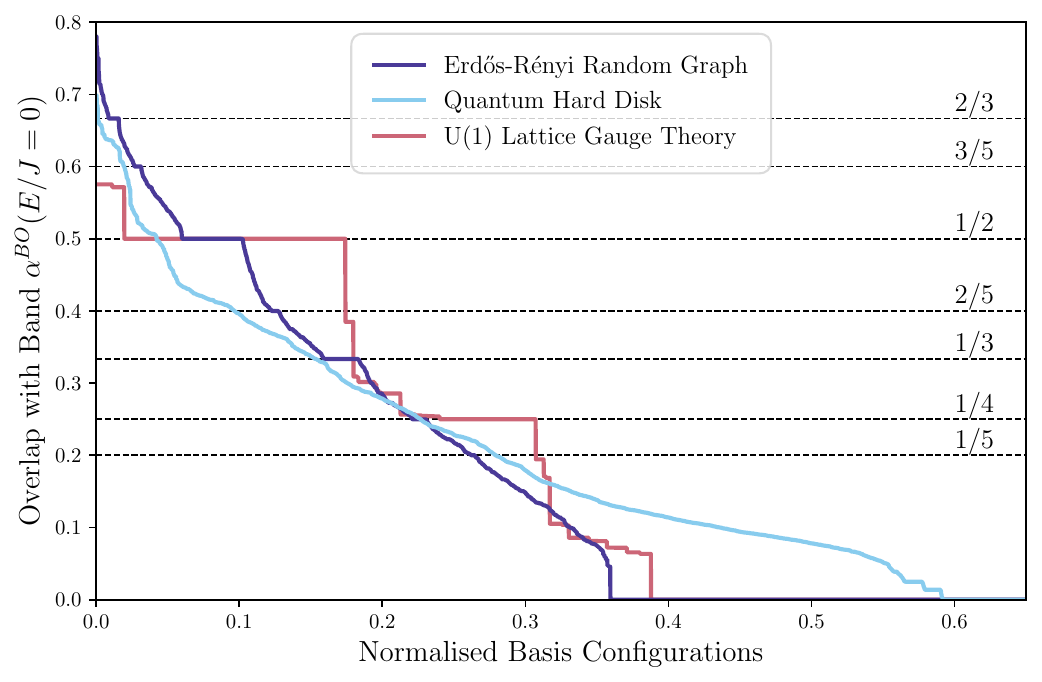}\:
\caption{\textbf{The band-overlap order parameter $\qea(\ell,0)$ for the $\epsilon=0$ flat band.}
States $\ket{\phi_\ell}$ are ordered by decreasing $\qea(\ell, 0)$. The abscissa shows the corresponding rank index normalised by the Hilbert-space dimension.
The curves are shown for the three models  in Fig.~\ref{fig:spec}: Erd\H{o}s-R\'{e}nyi graph, quantum hard-disk model, and a $U(1)$ lattice gauge theory. The fifth-order Farey sequence is shown in black dashed lines.}
\label{fig:devilsS}
\end{figure}

\textit{Many-body-caged spin glass --}
Many-body cages thus yield quantum states inaccessible in thermal equilibrium.
This is encoded in what we call a many-body-caged spin glass, characterised by the following properties: i) it occurs in the absence of disorder, even in $D>1$; ii) the flat bands comprise a finite fraction of Hilbert space. We therefore identify this as eigenspectrum order---a variety of eigenstate order~\cite{nandkishore2015many}---resulting from many-body caging.
This differs from conventional eigenstate order in that it is hosted exclusively on the flat bands of the many-body spectrum.

In systems that exhibit eigenstate order, the order parameter reveals nontrivial structure in a large fraction of eigenstates, including at high energies where such structure would be absent in thermalising systems~\cite{nandkishore2015many}.
In homogeneous systems, many-body cages form a degenerate manifold. 
We therefore construct an order parameter $\qty(\qea)$ that characterises the structure of this manifold in the many-body basis.
Specifically, we evaluate $\qea$ through the overlap between the many-body basis states and the eigenstates comprising the flat band at energy $\varepsilon$.
In physical systems, the band-overlap order parameter represents how much the system and its local observables remain non-thermal due to many-body cages (See SM for more on the hard-disk model).
For a given basis state $\ket{\phi_{\ell}}$ and energy $\varepsilon$, we define:
\begin{equation} \label{eq:EA_Fock}
    \qea(\ell, \varepsilon) = \sum_{n|E_n = \varepsilon} \abs{\braket{\phi_\ell}{E_n}}^2,
\end{equation}
where $\ket{E_n}$ are energy basis eigenstates.
This quantity is independent of the explicit choice of eigenbasis in the flat bands (for the two systems we consider, we use the Fock basis).

In systems governed by ETH, $\qea(\ell, \varepsilon)$ decays exponentially with system size~\cite{d2016quantum}. In the many-body-caged spin glass, $\qea(\ell, \varepsilon)$ remains nonzero. 
Evaluating $\qea(\ell, \varepsilon)$ across all basis states and ordering by the magnitude of $\qea$ reveals rich structure (Fig.~\ref{fig:devilsS}). 

The $\qea$ curves characterise the localisation of many-body cages via their overlap with individual basis states; because these cages are compact, i.e., no exponential tails, the resulting curves encode their underlying motifs.
Specifically, those supported on strictly local motifs (e.g. tree structures) often give rise to rational values of the $\qea(\ell, \varepsilon)$, reflecting their symmetric support on a finite number of sites. Meanwhile, plateaus in $\qea(\ell, \varepsilon)$ indicate multiple similarly structured many-body cages, and a corresponding set of basis states exhibiting comparable long-time dynamics.
Thus, $\qea(\ell,\varepsilon)$ provides a probe of the structure within the degenerate subspaces.
We note that these curves can also exhibit non-integer fractal dimensions, discussed further in the End Matter.
\begin{figure}[!htb] 
\centering
\includegraphics[width=1\linewidth]{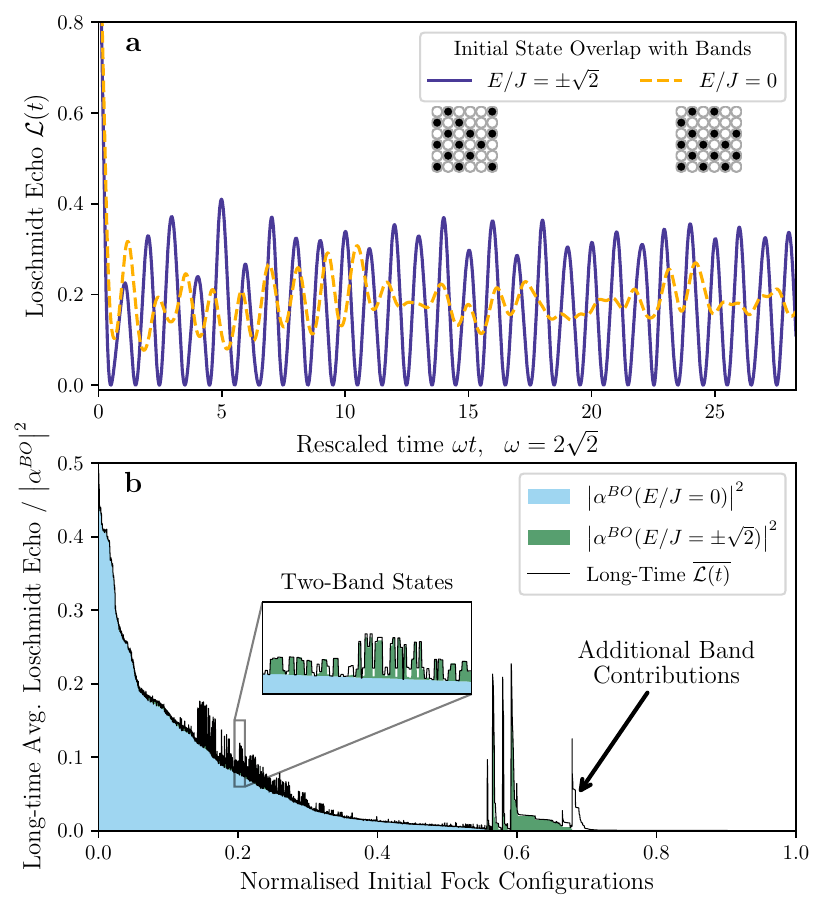}\:
\caption{\textbf{Dynamical signatures of the many-body-caged spin glass.} \textbf{a} Loschmidt echo $\mathcal{L}(t)$ in a $6\times6$ quantum hard-disk model for an initial Fock states overlapping a single ($E/J=0$) and multiple ($E/J=\pm\sqrt{2}$) flat bands. $\mathcal{L}(t)$ exhibits regular and long-lived many-body Rabi oscillations, absent in the single-band overlap case. \textbf{b} Statistics of the long-time behaviour of the Loschmidt echo in comparison to the band-overlap order parameters $\qea(\ell, \varepsilon)$ for the hard-disk system shown in Fig.\ref{fig:spec}b. 
Initial states $\ket{\phi_\ell}$ are ordered in descending order of $\qea(\ell, 0)$ (shaded light blue region) followed by $\qea(\ell, \sqrt{2})$ (shaded green region). 
The abscissa shows the corresponding rank index normalised by the Hilbert-space dimension.
Initial Fock states with further flat band overlaps (e.g. $E/J=\pm\sqrt{3}$) correspond to the indicated additional band contributions. }
\label{fig:mbro_le}
\end{figure}

\textit{Many-body Rabi oscillations -- }
We next explore the spatiotemporal aspect of this eigenspectrum order. 
In particular, we show that the flat bands give rise to coherent many-body Rabi oscillations at characteristic frequencies(see Fig.~\ref{fig:mbro_le}) which can be detected via, e.g., Loschmidt echoes~\cite{karch2025probing, lunkin2026evidence}) demonstrates. We define the Loschmidt echo as,
\begin{equation}
    \mathcal{L}(t) = \abs{\braket{\psi}{\psi(t)}}^2\ .
\end{equation}
Here $\ket{\psi}$ denotes the initial basis state, and $\ket{\psi(t)} \equiv e^{-iHt} \ket{\psi}$ is the time-evolved state with $H$ the underlying microscopic Hamiltonian.
For quantum systems satisfying ETH, $\mathcal{L}(t)$ generally becomes exponentially small in system size at late times~\cite{d2016quantum}.

The key dynamical signature of many-body-caged spin glasses are long-lived oscillations
in $\mathcal{L}(t)$, with a finite amplitude that is independent of system size.
Fig.~\ref{fig:mbro_le}a shows such many-body Rabi oscillations for an initial state overlapping two flat bands, in a $6\times6$ hard-disk model: 
the dynamics takes place in an effectively few-level system (see End Matter) with energies set by the flat-band eigenvalues. This explains the appearance of characteristic frequencies for the case of flat bands due to tree grafting, as well as the absence of these oscillations for a state overlapping a single flat band. Analogous oscillations appear for the hard-disk autocorrelation function, see the SM.

Fig.~\ref{fig:mbro_le}b shows the long-time-averaged Loschmidt echo $\overline{\mathcal{L}} = T^{-1} \int_t^{t+T} dt' \mathcal{L}(t')$, for all initial basis states for the hard disk model at $Jt=10^9,\, T=100$.
We compare this to the overlaps $\qea(\epsilon)$ with the dominant flat bands at $\epsilon/J=0,\pm \sqrt{2}$ for the system size $5\times8$ considered in Fig.~\ref{fig:spec}.

Overall, the dynamics associated with many-body cages is governed by an effectively few-level system, leading to persistent, system-size–independent oscillations in the Loschmidt echo and related dynamical correlation functions (see SM). The long-time dynamics retain initial-state memory, directly reflected in $\bar{\mathcal{L}}$. This residual signal provides a probe of $\qea(\epsilon)$ from dynamical simulations.

\textit{Discussion \& Outlook -- }
A central aspect of this work is its construction of a theoretical framework of many-body cages and flat bands on the state graph, which interprets and connects non-equilibrium phenomena observed in multiple different models.
In this context, several previous works showcase phenomena that naturally fit this framework. Works on the PXP model \citep{brighi2023hilbert, ganguli2025aspects, turner2018weak, surace2023quantum} note the existence of zero-energy states and investigate zero/low-entanglement states in the context of scars. In the single-particle case, quantum percolation \citep{kirkpatrick1972localized, harris1982exact} identifies a similar localisation mechanism in real space, while more recent work \citep{stern2021quantum} transposes this to quantum spin ice, which we hypothesize also involves many-body caging. Works on quantum combs identify CLS in real-space tree structures~\cite{hart2020compact}, while many-body cages occur purely on the state-graph.

Many-body cages represent a distinct phenomenon from previously discovered mechanisms for nonequilibrium phenomena such as Hilbert-space fragmentation (HSF) and disorder free localisation (DFL).
HSF arises when kinetically or energetically \textit{disconnected} clusters emerge in the state graph, leading to multiple disconnected Krylov subspaces that prevent the system from thermalising~\cite{moudgalya2022quantum, sala2020ergodicity}. In contrast, many-body cages are localised eigenstates that exist within the fully connected GCC and suppress thermalisation through interference, rather than via disconnected Krylov subspaces; while proximity to weak HSF is a favourable setting for their emergence, it is likely not a prerequisite.
DFL results from an effective disorder landscape generated by static background charges through the local gauge symmetries, while many-body cages result from local interference effects, independent of the complete system.
Finally, quantum many-body scars are described as measure-zero, \textit{isolated}, non-thermal eigenstates embedded in an otherwise thermal spectrum. Many-body cages, on the other hand, construct discrete flat bands in the spectrum, of potentially finite fraction of Hilbert space. The introduction of disorder or other perturbations could connect the two phenomena through a crossover or transition. This constitutes an interesting future direction, which, however, requires further work to establish rigorously.

The term many-body flat bands has been used for specific many-body states constructed from single-particle flat-band states(see, e.g.,\citep{danieli2021quantum, santos2020methods}). In our work, many-body flat bands emerge as a property of the many-body state graph, independent of a single-particle picture.

Our work not only presents a natural extension of flat-band physics to the many-body state graph, but also gives attainable experimental signatures across multiple platforms—such as Rydberg atoms acting as 2D hard-disk systems or ultracold atom quantum simulation platforms hosting lattice gauge theories \citep{wiese2013ultracold, aidelsburger2022cold}.

Future directions include exploring whether many-body flat bands exhibit a similarly rich zoology to that of single-particle flat band systems, such as topological (kagome \cite{1992PhRvL..68..855C}) or imbalanced bipartite (Lieb \cite{1989PhRvL..62.1201L} or diluted graphene \cite{2009RvMP...81..109C}) lattices.
Notably, since bipartite imbalance plays a substantial role in lower bounding the size of the $E=0$ band, it offers an additional route for introducing many-body cages into the state graph~(See Fig. S2b in the SM).
The high dimensionality of the state graph may offer additional features, as well as potential richer topological features \cite{bergman2008band,2023PhRvE.107b4125P}. 

Finally, single-particle (near-)flat bands arising from lattice geometry have long been known to enhance correlations~\cite{checkelsky2024flat}, and, e.g., promote superconductivity and other correlated states in Moiré systems~\cite{cao2018unconventional, bhowmik2024emergent}. In analogy, the statistical mechanics of a microcanonical ensemble, defined by projecting the many-body system onto a many-body flat band, is also a direction worth exploring: what--possibly exotic--properties, instabilities, or other phenomena may be discovered there?

\begin{acknowledgments}
\textit{Acknowledgments -- }
The authors thank Anushya Chandran, Cheryne Jonay, Vighnesh Dattatraya Naik, Frank Pollmann, Daniel Schmieg and Fabian Ballar Trigueros for valuable discussions and comments. This work was in part supported by the Deutsche Forschungsgemeinschaft under grants FOR 5522 (project-id 499180199) and the cluster of excellence ct.qmat (EXC 2147, project-id 390858490).
This project has received funding from the European Research Council (ERC) under the European Union’s Horizon 2020 research and innovation programme (grant agreement No. 853443).
The authors gratefully acknowledge the resources on the LiCCA HPC cluster of the University of Augsburg, co-funded by the Deutsche Forschungsgemeinschaft (DFG, German Research Foundation) – Project-ID 499211671.

{\bf Note added:} Three related pieces of work have appeared on the arXiv around the same time as our work \citep{tan2025interference, nicolau2025fragmentation, jonay2025localized}.

\end{acknowledgments}
\textit{Data availability —} The data to generate all figures in this letter is available in Zenodo \citep{data_repo}

\newpage

\appendix*
\section*{End Matter}

\newcounter{methodsection}
\renewcommand{\themethodsection}{\Roman{methodsection}}

\textit{Quantum Hard-Disk Model--- }
The quantum hard-disk model considered in this work is based on the model discussed in Ref. \cite{naik2024quantum}. The model Hamiltonian is:
\begin{equation}
    H = J\sum_{\left<i, j\right>} P_i \left( a_i^\dagger a_j + a_j^\dagger a_i \right) P_j,
\end{equation}
where $a_i, a_i^\dagger$ are the annihilation and creation operators of hard-core bosons respectively, and the projection operator $P_i$ enforces the Hard-disk constraint,
\begin{equation}
    P_i = \prod_{j\ni \abs{r_i - r_j}=1} \left( 1 - n_j \right) ,
\end{equation}
where $n_j \equiv a^\dagger_i a_i$, and the density of particles in the system is defined as $\rho(N, L) = \frac{N}{L_1L_2}$,
with N the number of particles, and $L_1,L_2$ the extent of the system. 

Long-time memory in the hard-disk model can be identified using the autocorrelation function, 
\begin{equation} \label{eq:HD_Ct}
C(t) = 
\frac{\frac{1}{L^2} \sum_{i} \expval{\qty(2 \hat{n}_i(t) - 1)\qty(2 \hat{n}_i(0) - 1)} - C^*}{1-C^*}
\end{equation}
where $C^* = \left(2\rho - 1\right)^2$ ensures that $C(t) \rightarrow 0$ if no long-time memory exists. 

As in \cite{naik2024quantum} we consider a square lattice, and enforce a nearest neighbour hard-disk constraint. Finally, all hard-disk systems in this article have closed boundary conditions.

\textit{U(1) Quantum Link Model--- }
The U(1) lattice gauge theory used in this work is based on the spin-1/2 quantum link model discussed by \cite{karpov2021disorder} and modified by \cite{chakraborty2025fractional}, named `Stochastic dynamical charge hopping model'. We implement a quantum analogue of it.
Choosing a random configuration of background charges on the quantum link model, we then consider the hopping problem of a single dynamical charge, while conserving the gauss' law constraint. 
The model Hamiltonian is:
\begin{equation}
H = J \sum_{\vb{r}, \vu*{\mu}} \qty( \sigma^-_{\vb{r}} S^+_{\vb{r}+\vu*{\mu}} \sigma^+_{\vb{r}+\vu*{\mu}} + h.c.),    
\end{equation}
where $S^{+(-)}$ are the spin raising(lowering) operator representing the gauge spins on the links, $\sigma^{+(-)}$ are the spin raising(lowering) operator representing the dynamical particle on the vertices, $\vb{r} = (x,y)$ is the vertex position, and $\vu*{\mu} =(\vu*{i},\vu*{j})$ is one of the two unit vectors of the lattice.

\textit{Fractality in $\qea$ Curves--- }
In the Erd\H{o}s-R\'{e}nyi random graph (Fig.~\ref{fig:devilsS}, dark blue), we observe clear plateaux, occurring at values of $\qea$ corresponding to the Farey sequence of order $n=5$ (the sequence of all completely reduced fractions between $0$ and $1$ whose denominators are less than or equal to $n$; See SM for system-size scaling). 
Specifically, $\qea(\varepsilon)$ for a flat band can exhibit a fractal structure; in the Erd\H{o}s-R\'{e}nyi graph this manifests as a devil's staircase.

We characterise this fractal behaviour by determining the set where the $\qea$ curve varies, computing its fractal dimension $D_f$ via box-counting~\cite{klinkenberg1994review} (see SM for details).
We present bounds on $D_f$---rather than exact values---due to uncertainties inherent to the box-counting technique.

For the Erd\H{o}s-R\'{e}nyi graph, we find $D_f \leq 0.917 < 1$, signalling a fractal $\qea$ dimension consistent with a devil's staircase.
For the hard-disk system (Fig.~\ref{fig:devilsS}, light blue) we find no clear signature of fractality ($D_f\approx 1$). The lattice gauge theory (Fig.~\ref{fig:devilsS}, light red) displays clear plateaux, suggesting a fractal dimension of $D_f\leq 0.75$.

\textit{Derivation of Many-Body Rabi Oscillations--- }
Since Fock states of the hard-disk model overlap bands symmetrically, we consider a state with substantial overlap in the $\pm E$ bands of a generic system,
\begin{equation} \label{eq:mbstate1}
    \ket{\psi(t)} = \sum_{n}\alpha_ne^{iE_n t} \ket{-E_n} + \sum_{n}\alpha_ne^{-iE_n t} \ket{E_n} + \mathrm{nonflat},
\end{equation}
where $\ket{E_n}$ is an arbitrary basis in the degenerate band. Since inside the degenerate band a basis may be chosen where the overlap of the basis state, $\ket{\psi}$, is on a single eigenstate, this can be rewritten as,
\begin{equation} \label{eq:mbstate2}
    \ket{\psi(t)} = ae^{iEt} \ket{\phi_{-E}} + ae^{-iEt} \ket{\phi_{+E}} +  \mathrm{nonflat},
\end{equation}
where $a = \sum_n\alpha_n$, and $\phi_{\pm E}$ are the states of the rotated eigenbasis. Assuming the overlap outside the bands is small, with exponentially decaying fluctuations, the overlap evaluates to,
\begin{equation}
    \braket{\psi(0)}{\psi(t)} = 2a^2 \cos(Et),
\end{equation}
and the Loschmidt echo
\begin{equation}
    \mathcal{L}(t) = 4|a|^4\cos^2(Et). 
\end{equation}

\textit{Long-Time Memory--- }
The relation between the band-overlap order parameter, $\qea(\psi, \varepsilon)$, and the Loschmidt echo can be shown analytically. We apply the same procedure as in Eq.~\ref{eq:mbstate2} to an initial basis state that overlaps with three bands, $\varepsilon=0, \pm E$. The state can be written as
\begin{equation}
        \ket{\psi(t)} = ae^{iEt} \ket{\phi_{-E}} + b \ket{\phi_{0}} + ae^{-iEt} \ket{\phi_{+E}} + \mathrm{nonflat}).
\end{equation}
The Loschmidt echo then evaluates to,
\begin{equation}
    \mathcal{L}(t) = 4|a|^4 \cos^2(Et) + 2|a|^2|b|^2\cos(Et) + |b|^4,
\end{equation}
giving the time-averaged quantity,
\begin{equation}
    \bar{\mathcal{L}} \approx 2|a|^4 + |b|^4
\end{equation}
The band-overlap order parameters, defined in Eq.~\ref{eq:EA_Fock}, correspond to the overlap of the initial basis state with each band independently. Squaring and summing over all bands yields,
\begin{equation}
    \sum_{\varepsilon = 0, \pm E}|\qea(\ell, \varepsilon)|^2 = 2|a|^4 + |b|^4,
\end{equation}
showing that the two quantities are in agreement. 

Furthermore, in the hard-disk model we look at the long-time memory through local observables, using the autocorrelation function in Eq. \ref{eq:HD_Ct}.
\begin{figure}[!hbt] 
\centering
\includegraphics[width=1.0\linewidth]{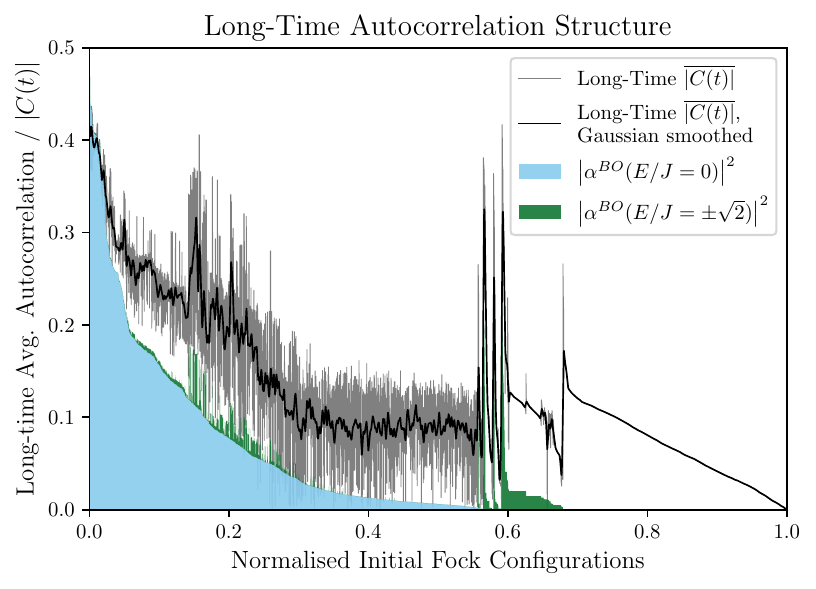}\:
\caption{
Statistics of the long-time behaviour of the autocorrelation function in comparison with the band-overlap order parameters $\qea$ for the hard-disk model in Fig.~\ref{fig:spec}b. Fock configuration numbers (abscissa) are divided by the Hilbert space dimension, and are ordered with descending overlap with the $E/J=0$ flat band, followed by the $E/J=\pm \sqrt{2}$ flat band overlap. The black curve shows the Gaussian smoothed $|C(t)|$ to highlight the overall trend as $\qea$ decreases.
}
\label{fig:HD_C}
\end{figure}

\textit{Band Robustness--- }
Overall, perturbations of the Hamiltonian can be distinguished into two different classes.

\begin{figure}[!hbt] 
\centering
\includegraphics[width=1.0\linewidth]{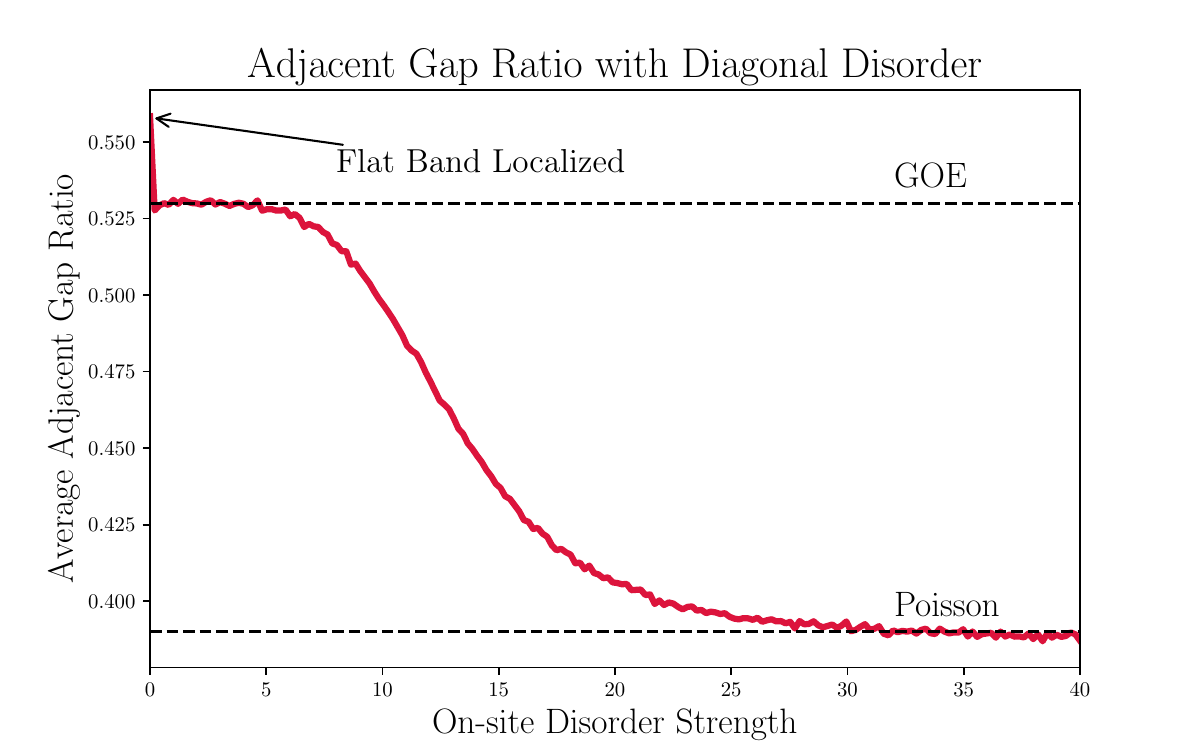}\:
\caption{\textbf{Reentrant `Anderson' delocalisation.} 
Introducing uncorrelated on-site disorder in the sparse Erd\H{o}s-R\'{e}nyi graph breaks the many-body cages and destroys the spin glass, recovering GOE level statistics. At higher disorder strength the system undergoes Anderson localisation reflected in Poissonian level statistics.}
\label{fig:A_Loc}
\end{figure}

Firstly, those with diagonal terms in the state graph, which can e.g.\ emerge from local potentials in the underlying microscopic Hamiltonian.
Introducing an uncorrelated disorder potential into the Erd\H{o}s-R\'{e}nyi random graph reveals a reentrant `Anderson' delocalisation, which destroys the CLS on the many-body cages and thus the flat bands. This is followed by an MBL phase, to the extent that it exists, as shown in Fig.~\ref{fig:A_Loc} for the Erd\H{o}s-R\'{e}nyi model. 

A different kind of diagonal terms can emerge from interactions in the microscopic Hamiltonian.
These diagonal terms are not entirely random, but rather realize a correlated disorder. For judicious choices,  this can preserve at least some of the bands. 
As a simple pedagogical example, if the diagonal terms of a grafted tree on the state graph all have the same on-site values, the many-body cage is preserved, albeit the corresponding eigenvalue will exhibit a shift.
For more general soft-core interactions, numerical simulations done on the quantum hard disk model \citep{trigueros2025dynamics} provide evidence that some of the many-body cages can indeed survive.
Thus, systems with physically realizable soft-core interactions could still preserve and exhibit our phenomenology. 

A second type of perturbation is kinetic disorder, represented as off-diagonal disorder on the state graph. This does not destroy the $E/J=0$ CLS, but does affect other bands. 
Since the $E/J=0$ is the dominant band, and the backbone of the non-thermal behaviour of the system, the long-time memory of the system will remain. However, the richness that comes from the interplay of multiple bands can be potentially suppressed.
We include detailed calculations of these effects in the SM section III. 

\bibliography{references}

\onecolumngrid

\pagebreak

\appendix
\section*{Supplementary Material: \\ Many-body cages: disorder-free glassiness from \\ flat bands in Fock space, and many-body Rabi oscillations }

\title{Supplementary Material \\ Many-body cages: disorder-free glassiness from \\ flat bands in Fock space, and many-body Rabi oscillations }

\date\today

\author{Tom Ben-Ami}
\affiliation{Theoretical Physics III, Center for Electronic Correlations and Magnetism, Institute of Physics, University of Augsburg, D-86135 Augsburg, Germany}
\affiliation{Max-Planck-Institut f\"{u}r Physik komplexer Systeme, N\"{o}thnitzer Stra\ss e 38, Dresden 01187, Germany}

\author{Markus Heyl}
\affiliation{Theoretical Physics III, Center for Electronic Correlations and Magnetism, Institute of Physics, University of Augsburg, D-86135 Augsburg, Germany}
\affiliation{Centre for Advanced Analytics and Predictive Sciences (CAAPS), University of Augsburg, Universitätsstr. 12a, 86159 Augsburg, Germany}

\author{Roderich Moessner}
\affiliation{Max-Planck-Institut f\"{u}r Physik komplexer Systeme, N\"{o}thnitzer Stra\ss e 38, Dresden 01187, Germany}

\maketitle

\section{Band-Size Finite Size Analysis} \label{sec:fsa}
\begin{figure}[htbp]
    \centering
    \includegraphics[width=0.6\textwidth]{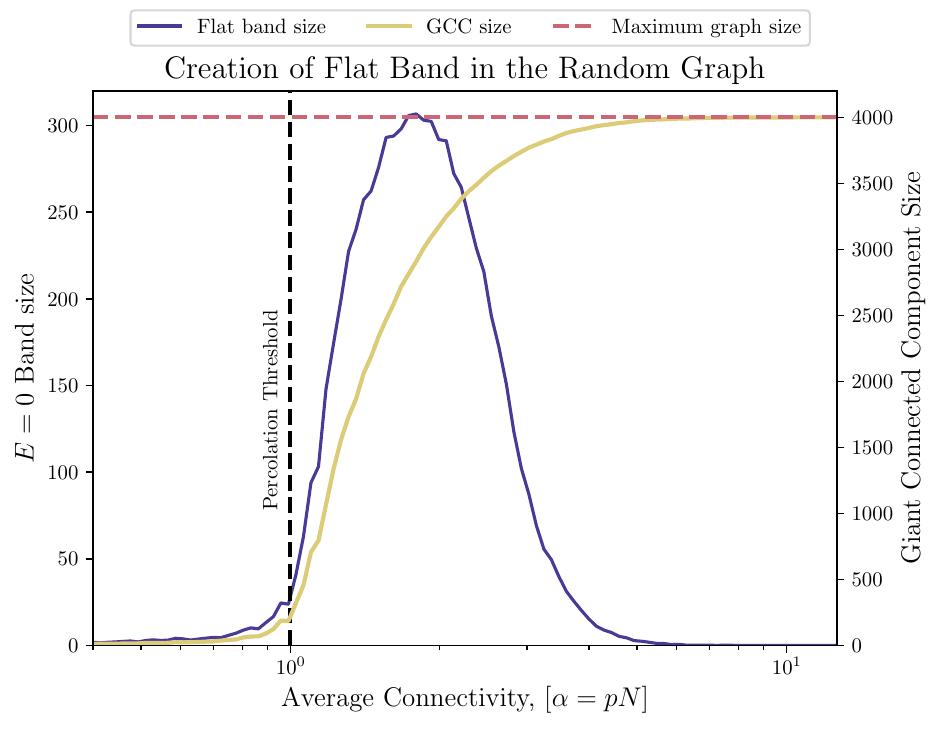}
    \caption{\textbf{Emergence and disappearance of the $E=0$ flat band}. The size of the $E=0$ flat band is plotted as a function of average connectivity in the Erd\H{o}s-Reny\'{i} random graph [right axis] alongside the size of the giant connected component [left axis]. Data were averaged over 20 graph realisations of $N=4000$ nodes. The black vertical line represents the expected theoretical percolation transition, $pN=1$, beyond which no giant connected component exists.
    }
    \label{fig:RG_creation}
\end{figure}

To determine the prominence of flat bands, we focus on how the size of the $E=0$ band scales with  Hilbert space dimension. We use random graphs, which provide a highly tunable setting where both the sparsity (analogous to the constraint strength), and the graph size (analogous to the Hilbert space dimension) can be controlled. 
In Fig.~\ref{fig:RG_creation} we associate the emergence of the $E=0$ band in the Erd\H{o}s-Reny\'{i} random graph with increased sparsity, and its disappearance at the percolation transition, where the band size is limited by the vanishing size of the giant connected component (GCC), the largest connected component in the graph.

Since Fig.~\ref{fig:RG_creation} shows a wide range of $p$ where the flat band is prominent, we concentrate in the following analysis on the peak of the curve where the $E=0$ flat band is largest. We implement the following optimization algorithm for finding the
edge probability that gives the largest $E=0$ band, $p$:
\begin{enumerate}
    \item[Step 1] \textbf{Coarse scan:}
    \begin{enumerate}
        \item Evaluate the band size at eight logarithmically spaced edge probabilities across the range of $p$. 
        \item Identify the edge probability with the largest band size, $p_i$.
        \item Narrow the search window to the logarithmic interval between the neighbouring points around $p_i$.
    \end{enumerate}
    \item[Step 2] \textbf{Golden-section search:}
    \begin{enumerate}
        \item Compute the band size at two interior points, $[p_1, p_2]$, within the current interval using the golden ratio.
        \item Retain the subinterval corresponding to the larger band size and discard the other.
        \item Continue until the interval width is less than a specified tolerance, yielding $p_{gs}$.
    \end{enumerate}
    \item[Step 3] \textbf{Fine-tuning:}
    \begin{enumerate}
        \item Sample 10 values of $p$ uniformly spaced in logarithmic scale within a window of width 0.1 (log-scale) centred around $p_{gs}$.
        \item Evaluate the band size at each point and update the final best estimate.
    \end{enumerate}
\end{enumerate}

\subsection{Random Graphs} \label{subsec:rg_fsa}   

\begin{figure}[htbp]
    \centering
    \includegraphics[width=1.0\textwidth]{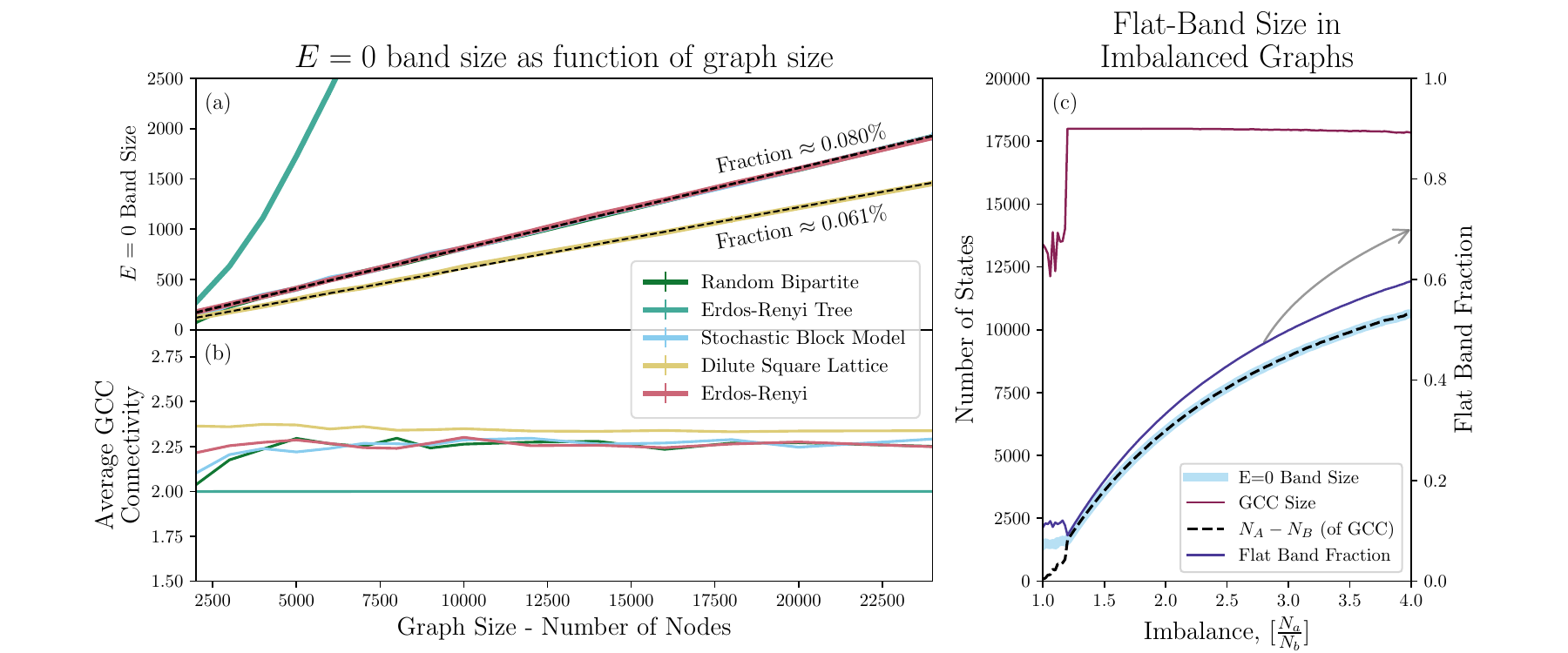}
    \caption{\textbf{Finite size analysis of sparse random graphs}. \textbf{a} Size of the $E=0$ band as a function of graph size for different types of random graphs. The band sizes of the random bipartite, stochastic block model, and Erd\H{o}s-Reny\'{i} graphs overlap. \textbf{b} Average connectivity of the giant connected component as a function of graph size. \textbf{c} Maximum size of $E=0$ flat band as a function of imbalance, $N_a/N_b$, in the random bipartite graph. For \textbf{a}\&\textbf{b} all measurements are averaged over 10 graph realisations, with the errorbars representing the standard deviation. \textbf{c} is measured over a single graph realization.}
    \label{fig:RG_fsa}
\end{figure}

While in the Erd\H{o}s-Reny\'{i} random graph, where edges are completely uncorrelated, meaning a graph ensemble with maximal entropy,
 we consider how different graph ensembles affect the existence of many-body cages and flat bands. In addition, we pinpoint the ideal sparsity of the giant connected cluster for the emergence of flat bands in random graphs. The results are summarised in Fig.~\ref{fig:RG_fsa}.

\paragraph{Bipartiteness}
We impose bipartiteness on the Erd\H{o}s-Reny\'{i} graph.
Bipartiteness is common in the state graphs of physical systems, for example in the square lattice hard-disk model. 
Since trees are inherently bipartite, this restriction does not affect the existence of compact localized states and many-body flat bands. In Fig.~\ref{fig:RG_fsa}, the size of the flat bands in the \textit{balanced} bipartite case grows at exactly the same rate as in the Erd\H{o}s-Reny\'{i} case, strengthening the claim that this restriction does not affect the existence of dangling trees.

We also independently tune the random bipartite imbalance (Fig.~\ref{fig:RG_fsa}(c)). At increasing imbalance, we see that the maximum attainable flat band size also increases. Thus, many-body systems where the state graph has an imbalanced bipartite structure are good candidates for finding many-body cages. 

\paragraph{Loops} 
While Erd\H{o}s-Reny\'{i} graphs are locally tree-like, we isolate the effect of loops by generating a `random-connectivity tree' - a spanning tree of an Erd\H{o}s-Reny\'{i} graph. 
By removing loops, the spanning tree is saturated with dangling trees, and the size of the $E=0$ flat band grows rapidly. 
We note that since the optimization algorithm fine-tunes the system to find the largest $E=0$ band size, it is likely that the resultant spanning trees found are ones which strongly enhance the $E=0$ band specifically. 

\paragraph{Underlying Structure}
We can alternatively construct a sparse random graph by randomly diluting a dense graph, similar to a real-space concept introduced in \cite{kirkpatrick1972localized}.

For example, starting from a square lattice graph, links are removed randomly with probability $1-p$. At adequately low values of $p$ a sparse giant connected component emerges. 
Although the fraction of states in the $E=0$ band is smaller, the size of the band still scales linearly with the Hilbert space dimension (Fig.~\ref{fig:RG_fsa}(a)). 

This approach can also be conceptually extended to random-regular graphs, where a small distinction lies in that even in the dense limit graphs look locally tree-like. 

\paragraph{Community Structure}
To introduce weak community structure into the random graph we use the stochastic block model (SBM) with the block probability matrix
\begin{equation}
    \mathcal{P}(p) = \mqty[ p & p/3 & p/4 \\
                p/3 & p & p/2 \\
                p/4 & p/2 & p
            ],
\end{equation}
where $p$ is tuned to find the sparse limit of the graph.
As long as the probability matrix allows for the existence of sparse giant connected clusters, we expect the behaviour to resemble that of Erd\H{o}s-Reny\'{i} graphs. 

\paragraph{Random Geometric Graphs}
A counterexample to the graph models discussed above is the random geometric graph. These graphs rapidly fragment from dense regions, which do not contain many-body cages, to small disconnected clusters.
Since no sparse giant connected clusters appear, flat bands also do not emerge. 

\subsection{Quantum Hard Disks} \label{subsec:qhd_fsa}

\begin{figure}[htbp]
    \centering
    \includegraphics[width=0.9\textwidth]{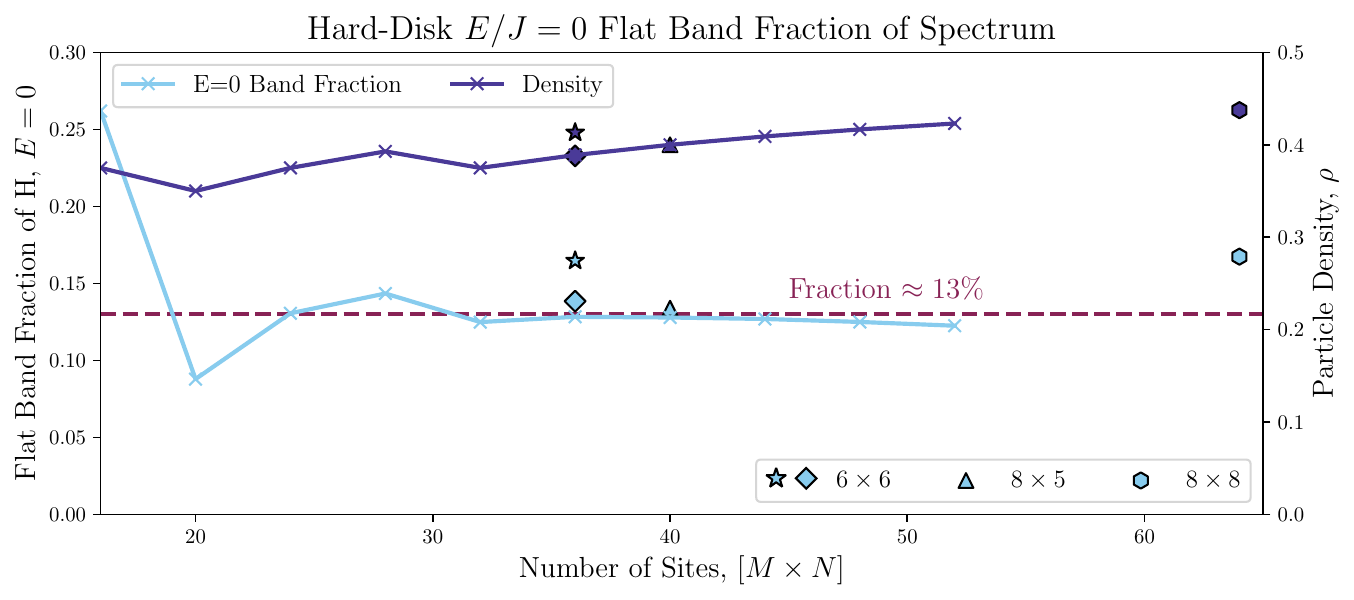}
    \caption{\textbf{Finite size analysis of hard-disk flat band size}.
    The fraction of the spectrum within the flat band is measured as a function of the number of sites, while keeping the density as constant as possible, $0.375 < \rho < 0.425$. The systems used are $4\times N$ strips (data points joined by lines), and only the largest Hilbert space fragment is analysed. 
    Additional markers represent system sizes of different sizes and aspect ratios.}
    \label{fig:QHD_fsa}
\end{figure}

In the hard-disk model the strength of the constraint is tuned by the particle density, $\rho$. Unlike in the random graphs where we can keep $p$ constant, for hard-disks $\rho$ cannot be kept at the same value 
as we increase the system size for numerically accessible system sizes. 
Thus, we consider a $4\times N$ system where we keep the density between $0.375$ and $0.425$, within the weak fragmentation regime and near the strong-fragmentation transition point \cite{naik2024quantum}. The results are shown in Fig.~\ref{fig:QHD_fsa}.

\section{Band-Overlap Order Parameter in Random Graphs} \label{sec:devilS}

\begin{figure}[htbp]
    \centering
    \includegraphics[width=1.0\textwidth]{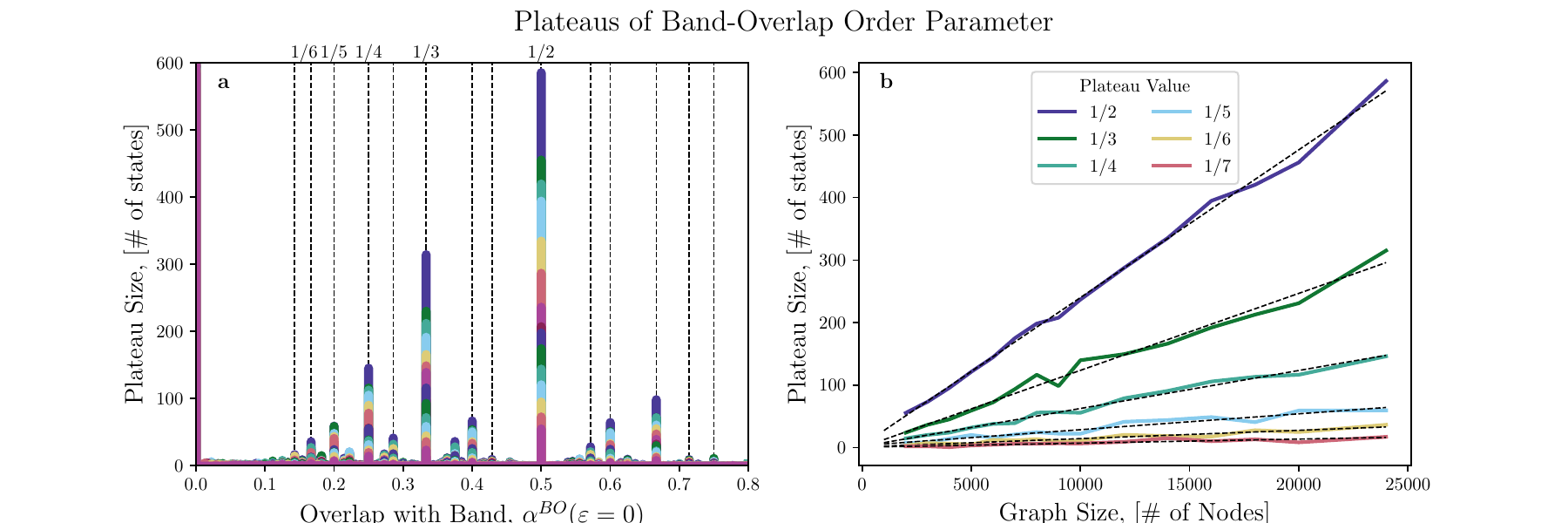}
    \caption{\textbf{Emergence of the devil's staircase in the Erd\H{o}s-Reny\'{i} random graph}. \textbf{a} Plateau widths of the Band-overlap order parameter. Colours indicate graph size, increasing from front to back. As the graph size is increased, more plateaux appear due to an increase in `resolution'. \textbf{b} Finite size analysis for the plateaux corresponding to the $1/n$ series of Farey fractions. Each data-point is averaged over 10 random graph realisations.}
    \label{fig:plateau_fsa}
\end{figure}

To identify the devil's staircase in Erd\H{o}s-Reny\'{i} random graphs, we analyse the width of the plateaux at Farey fractions as a function of the Hilbert space dimension. 
Fig.~\ref{fig:plateau_fsa} shows that for computationally accessible graph sizes, plateaux grow linearly with Hilbert space dimension, while additional new plateaux appear.  
The scaling behaviour coupled with the progressive appearance of additional plateaux strengthens the interpretation that the structure is indeed a devil's staircase.

\section{Robustness of Cages} \label{sec:robust}

\begin{figure}[!htbp]
    \centering
    \includegraphics[width=0.9\textwidth]{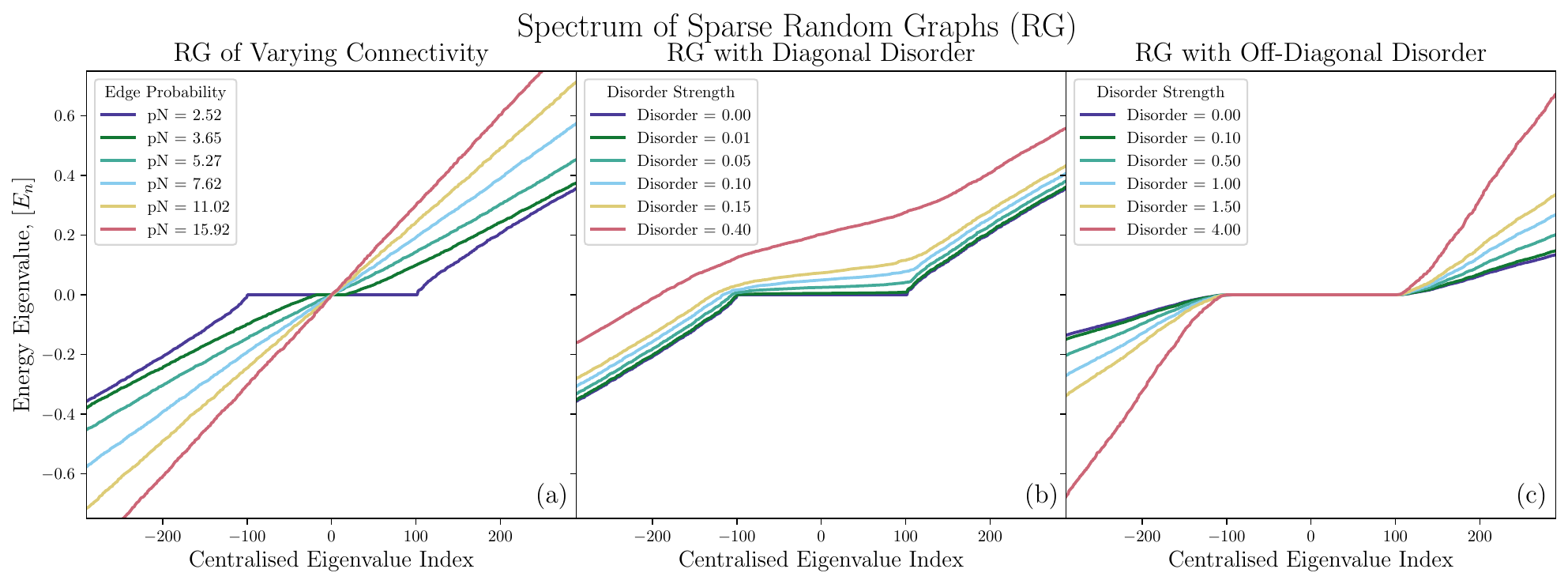}
    \caption{\textbf{Flat band formation \& robustness to disorder in random graphs}. \textbf{a} Emergence of the $E=0$ flat bands in the Erd\H{o}s-Reny\'{i} graph. \textbf{b} Destruction of the flat band under uncorrelated diagonal disorder. \textbf{c} Robustness of the $E=0$ flat band under uncorrelated off-diagonal disorder. All results are obtained from random graphs of 4000 nodes.
    }
    \label{fig:RG_rob}
\end{figure}

The robustness of the many-body cages is influenced by long-range interactions and kinetic disorder. 
While these introduce correlated disorder into the state graph, the nature of the correlations is highly system dependent.
Here, we test as a limiting case the stability of the bands under {\it uncorrelated} diagonal and off-diagonal disorder (Fig.~\ref{fig:RG_rob}). 

\section{Hard-disk Edwards-Anderson Order Parameter}
In the hard-disk model nodes directly correlate to many-body basis states. We can thus connect the long-time memory that arises from a single flat band in the case of a non-zero band-overlap order parameter to a non-zero Edwards-Anderson order parameter.
We first map the site occupation to a spin observable in the Heisenberg picture via $\hat{s}_i(t) = 2 \hat{n}_i(t) - 1$.
We additionally introduce the projector onto the degenerate many-body flat bands as $\hat{P}_\varepsilon = \sum_{n | E_n=\varepsilon} \ket{E_n}\bra{E_n}$, where $\ket{E_n}$ are energy eigenstates of the many-body spectrum.
Following the usual definition of the Edwards-Anderson order parameter from long-time autocorrelation:
\begin{equation}
    q_{EA} = \lim_{t\rightarrow \infty} \frac{1}{N} \sum_i \expval{\hat{s}_i(0)\hat{s}_i(t)},
\end{equation}
we consider the effect of many-body flat band on the autocorrelation of the hard disk model, introduced in the End Matter.
Since long-time memory arises solely from the weight of the initial basis state with the many-body flat band, we first project basis states onto the flat band as:
\[ \ket{\phi^\varepsilon_{\ell}} = \hat{P}_\varepsilon\ket{\phi_\ell},\]
with $w_\ell = \expval{\hat{P}_\varepsilon}{\phi_\ell}$ the spectral weight of the basis state in the band. The autocorrelation function may then be decomposed as,
\begin{equation}
    C_\ell(t) = C_{\ell}^{band}(t) + C_{\ell}^{thermal}(t)
\end{equation}
where the flat-band projected term becomes in the long-time limit,
\begin{equation}
    \lim_{t\rightarrow\infty} C_\ell^{band} (t) = \mel{\phi_\ell^\varepsilon}{s_i^2}{\phi_\ell^\varepsilon}.
\end{equation}
This contribution to the order parameter survives, since the flatness prevents dephasing. We assume here that the rest of the spectrum is non-degenerate and exhibits generic many-body dephasing so that its contribution to the autocorrelation decays at long times. Finally, by setting $s_i^2 = 1$, we find that 
\begin{equation}
    C_\ell(\infty) = \frac{w_\ell - C^*}{1 - C^*}.  
\end{equation}
Thus, we can finally take an ensemble average the autocorrelations over all possible initial basis states to find an Edwards-Anderson order parameter.


\section{Comparison to Ising Spin-Glass} \label{subsec:ising_sg}
\begin{figure}[htbp]
    \centering
    \includegraphics[width=0.7\textwidth]{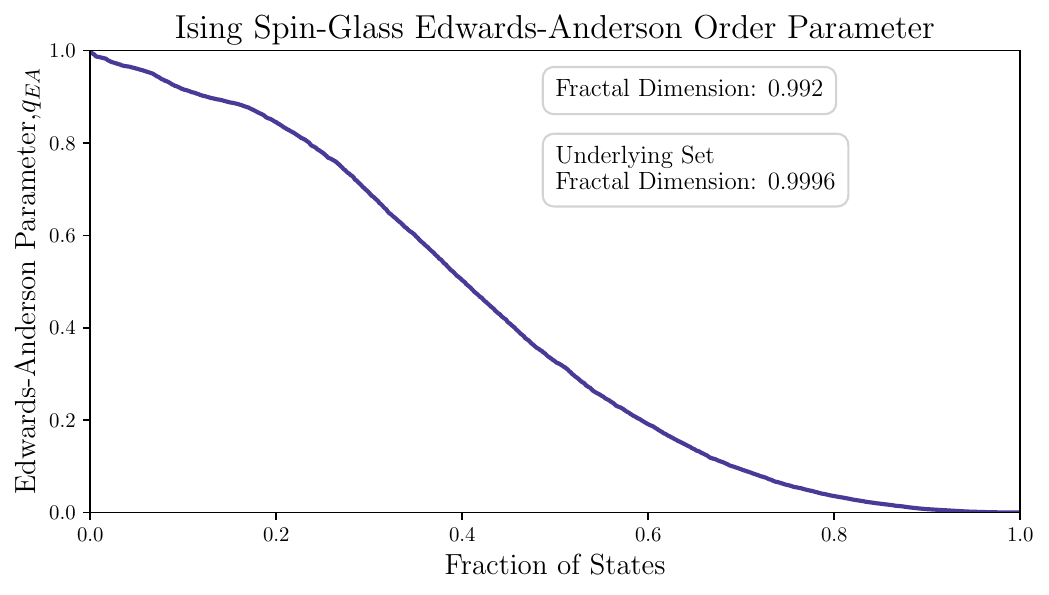}
    \caption{\textbf{Transverse-field Ising model Edwards-Anderson order parameter}. The Edwards-Anderson order parameter, arranged from largest to smallest, for a single realisation for the 1D TFIM of size $L = 14$, with $J \sim \mathcal{U}(-1.0, 1.0)$ and $h \sim \mathcal{U}(-\frac{J}{12} \frac{J}{12})$. }
    \label{fig:TFIM_EA}
\end{figure}

The Edwards-Anderson order parameter of the 1D transverse-field Ising model (TFIM) is defined as 
\begin{equation}
    q_{EA}^{\psi_n} = \frac{1}{L}\sum_{i=1}^{L}\abs{\mel{\psi_n}{S_i^z}{\psi_n}}^2,
\end{equation}
where $L$ is the system size, $\ket{\psi_n}$ are energy eigenstates of the system, and $S_i$ the spin at position $i$.
Fig.~\ref{fig:TFIM_EA} shows the structure of the TFIM Edwards-Anderson order parameter, demonstrating that the ordered $q_{EA}$ forms a smooth function with no fractal behaviour. 

\pagebreak

\section{Many-body Rabi Oscillations}
We look at the dynamics of the autocorrelation function of the hard-disk model. The resulting time-dependent dynamics shows rabi-oscillations of the same frequency seen in the Loschmidt echo plotted in figure 3.
Similarly, we find that for states which overlap only a single flat band, no oscillations are observed.

\begin{figure}[htbp]
    \centering
    \includegraphics[width=1.0\columnwidth]{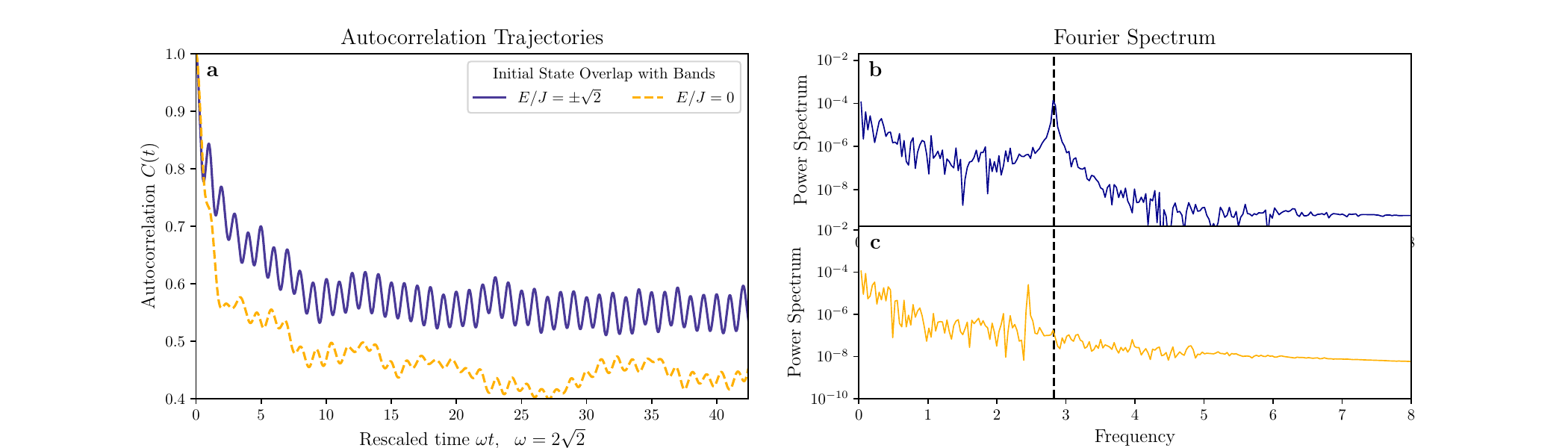}
    \caption{\textbf{Many-body Rabi oscillations in the QHD autocorrelation function}. \textbf{a} Time evolution of the autocorrelation function $C(t)$ for the initial basis states described in Figure 3\textbf{a}. \textbf{b}\&\textbf{c} the positive Fourier spectrum of the long-time dynamics for the two trajectories, showing the existence of persistent oscillations at $\omega=2\sqrt{2}$ when the initial state overlaps with two bands. 
    }
    \label{fig:MBRO_C}
\end{figure}

\section{Calculating the Fractal Dimension}
In the Erd\H{o}s-R\'{e}nyi random graph model, plateaux appear at values of Farey fractions (See Sec. \ref{sec:devilS}). This leads us to ask whether there exists an underlying fractal behaviour in the band-overlap order parameter. We turn to the devil's staircase - a continuous function that is constant almost everywhere, rising (or lowering) only on a measure-zero set of points. A famous example is the Cantor function.

Often, the underlying set of the staircase has unique fractal dimension, such as $D_f = \log(2)/\log(3)$ for the Cantor set. This underlying set can be deduced numerically from the staircase, $f(x)$, in finite-sized data by identifying where $f(x) - f(x+1) \neq 0$.

Thus, we investigate the fractal dimension of the underlying set of all three band-overlap order parameters in Fig.~2 using box-counting - a robust method for measuring fractal dimensions \cite{klinkenberg1994review}, which goes as follows:
\begin{enumerate}
    \item Span the space of the structure with boxes of size $\epsilon$.
    \item Count the number of boxes that contain the feature.
    \item Reduce the box size until its resolution matches the resolution of the data. 
\end{enumerate}
The fractal dimension is then estimated from the slope of the number of boxes as a function of box size.

Since finite sized data `smooths out' the underlying set as additional points are introduced to the set, we expect that measuring the fractal dimension will return a value higher than the actual value. Thus, our results likely give an upper bound.


\end{document}